\documentclass[a4paper,12pt,oneside,english]{article}
\usepackage{natbib} 
\usepackage{mathptmx}
\usepackage{amsmath}
\usepackage{amsfonts}
\usepackage{amssymb}
\usepackage{rotating}
\usepackage{subfigure}
\usepackage{graphicx}
\usepackage{babel}
\usepackage{epsfig} 
\usepackage{url}              % For breaking URLs easily trough lines
\newcommand{\aap}{   {\it Astron. Astrophys.}}

\newcommand{\apj}{    {\it Astrophys. J.}}
\newcommand{\apjl}{   {\it Astrophys. J. Lett.}}
\newcommand{\apjs}{   {\it Astrophys. J. Supplement.}}
\newcommand{\apss}{   {\it Astrophys. Space Sci.}}

\newcommand{\grl}{   {\it Geophys. Res. Lett.}}

\newcommand{\jgr}{    {\it J. Geophys. Res.}}
\newcommand{\mnras}{  {\it Mon. Not. Roy. Astron. Soc.}}
\newcommand{\nat}{    {\it Nature}}

\newcommand{\solphys}{{\it Solar Phys.}}

\newcommand{\ssr}{    {\it Space Sci. Rev.}}

\makeatother
\begin{document}
%%--------------Adding line numbers -------------------------
%\setpagewiselinenumbers
%\modulolinenumbers[1]
%\linenumbers
%%-----------------------------------------------------------
%
%\begin{article}
%
\noindent Accepted for publication in {\bf{\textit{Sol. Phys. - 2013}}}\\
\line(1,0){490}
\vspace{0.5cm}\\\\
{\LARGE{\bf{\noindent Changes in quasi-periodic variations of solar photospheric fields: precursor to the 
deep solar minimum in the cycle 23?}}}

\begin{center}
Susanta Kumar Bisoi${^1}$, P. Janardhan${^1}$, D. Chakrabarty${^2}$, \\ S. Ananthakrishnan${^3}$, and Ankur Divekar${^3}$
\end{center}
$\;\;$ \\
\noindent ${^1}$ {{Physical Research Laboratory, Ahmedabad - 380009, India. \\email: susanta@prl.res.in email: jerry@prl.res.in}} \\\\
\noindent $^{2}$ {{Physical Research Laboratory, Space \& Atmospheric Sciences Division, \\Ahmedabad - 380009, India. email: dipu@prl.res.in}} \\\\
\noindent ${^3}$ {{Electronic Science Department, Pune University, Pune 411 007, India. \\email: subra.anan@gmail.com e-mail: wiztronix@gmail.com}} \\

%%%%%%%%%%%%%%%%%%%%%%%%%%%%%%%%%%%%%%%%%%%%%%%%%%%%%%%%%%%%%%%%%%%%%%%%%%%%%%%%%%%%%%%%%%%%%%%%%%%%%%%%%%%%%%%%%%
\section*{abstract}

Using both wavelet and Fourier analysis, a study has been undertaken of the changes 
in the quasi-periodic variations in solar photospheric fields in the build-up to one 
of the deepest solar minima experienced in the past 100 years. This unusual and deep 
solar minimum occurred between solar cycles 23 and 24.  The study, carried out using 
ground based synoptic magnetograms spanning the period 1975.14 to 2009.86, covered 
solar cycles 21, 22 and 23.  A hemispheric asymmetry in periodicities of the photospheric 
fields was seen only at latitudes above $\pm$45${^{\circ}}$ when the data was divided, 
based on a wavelet analysis, into two parts: one prior to 1996 and the other after 1996.  
Furthermore, the hemispheric asymmetry was observed to be confined to the latitude range 
45${^{\circ}}$ to 60${^{\circ}}$.  This can be attributed to the variations in polar surges 
that primarily depend on both the emergence of surface magnetic flux and varying solar 
surface flows.  The observed asymmetry when coupled with the fact that both solar fields 
above $\pm$45${^{\circ}}$ and micro-turbulence levels in the inner-heliosphere have been 
decreasing since the early to mid nineties \citep{JaB11} suggests that around this time 
active changes occurred in the solar dynamo that governs the underlying basic processes 
in the sun.  These changes in turn probably initiated the build-up to the very deep solar 
minimum at the end of the cycle 23.  The decline in fields above $\pm$45${^{\circ}}$ for well 
over a solar cycle, would imply that weak polar fields have been generated in the past two 
successive solar cycles \textit{viz.} cycles 22 and 23.  A continuation of this declining 
trend beyond 22 years, if it occurs, will have serious implications on our current understanding 
of the solar dynamo.  
%------------------------------------------------------------------------------------------------
\section*{Introduction}
\label{S-Intro}
The delayed onset of solar cycle 24, after one of the deepest solar minimum experienced in the 
past 100 years has had significant solar and heliospheric consequences.  Solar cycle 23 has 
been characterized by a steady decline in solar activity \citep{McE08,JRL11}, a continuous 
weakening of polar fields \citep{JiC11} and a decline in micro-turbulence levels in the 
inner heliosphere since $\sim$1995 \citep{JaB11}.  Investigations of the boundary of polar 
coronal holes during the declining phase of solar cycle 23 using images from the Extreme 
Ultraviolet Imaging Telescope (EIT) on board the Solar Heliospheric Observatory (SoHO), have 
found a decrease in coronal hole area by $\sim$15$\%$ in comparison to that at the beginning of 
solar cycle 23 \citep{KiP09}.  Using both ground-based and space-borne observations of photospheric 
magnetic fields \citep{JBG10,HaR10} and theoretical modeling \citep{DiG10,NMM11}, there have been 
continuous efforts to investigate and understand the behavior of solar photospheric fields and 
their correlation with meridional flows, to try and explain the delayed onset of cycle 24 and 
the cause of the deep minimum at the end of cycle 23.  

Magnetic field measurements using data from the National Solar Observatory, Kitt Peak (NSO/KP) 
synoptic magnetogram database, have shown that a decline in solar photospheric fields in the 
latitude range 45${^{\circ}}$ to 78${^{\circ}}$ began around the minimum of solar cycle 22, 
in $\sim$1995$-$1996 \citep{JBG10}.  However, the dipole field of the Sun behaved differently.  
It was at its strongest in 1995, weakened at solar maximum around 2000, and then increased between 
$\sim$2000 and 2004 \citep{WRS09,JBG10}.  Signatures of this decline in solar fields above latitude 
$\pm$45${^{\circ}}$ have also been observed in the inner heliosphere as a corresponding decline 
in micro-turbulence levels which in turn, tie into small scale interplanetary magnetic fields 
\citep{ACK80,ABJ95}.  The decline in micro-turbulence levels was inferred from extensive 
interplanetary scintillation (IPS) observations at 327 MHz in the period 1983$-$2009 \citep{JaB11}.  
Using this fact {\it{viz.}} that both solar magnetic fields and micro-turbulence levels in the 
inner-heliosphere have been declining since mid 1990's, \citet{JaB11} have argued that the 
build-up to the deep and extended solar minimum at the end of cycle 23 was actually initiated as 
early as the mid-1990's.  

One method of gaining insights into underlying basic processes in the solar interior, that govern the 
nature and evolution of solar photospheric magnetic fields, is by studying periodicities produced by 
various surface activity features at different times in the solar cycle.  For example, the 158 days (d)
Rieger periodicity reported in the solar flare occurrence rates \citep{OBB98} has been linked to the 
periodic emergence of magnetic flux through the photosphere which in turn gives rise to a periodic 
variation of the total sunspot area on the solar surface.  Similarly, a strong 1.3 years (yr) periodicity 
detected at the base of the solar convection zone, through a helioseismic study \citep{HoC00}, has also 
been detected in sunspot areas and sunspot number time series studied using wavelet transforms \citep{KSo02}.  
These authors have in fact proposed that the 154${-}$158 d Rieger period is a harmonic of the 1.3 yr 
periodicity (3$\times$156 d = 1.28 yr) and that variations in the rotation rate have a strong 
influence on the workings of the solar dynamo \citep{KSo02}.

In this paper, in order to understand the role of periodic changes, if any, in the solar 
photospheric fields leading to the build-up of the recent deep minimum at the end of cycle 
23, data from the NSO/KP synoptic magnetic database during the period 1975.14~$-$~2009.86 
were subjected to detailed spectral analyses using both wavelet and Fourier techniques.  It 
has been shown that a significant north-south asymmetry in the quasi-periodic variations of the
photospheric magnetic fields exist both prior to and after 1996 (with 1996 being a clear 
transition period) almost 12 yr before the deep minimum period in 2008$-$2009.

\subsection*{Solar Periodicities}
\label{S-solar-periods}
Prominent periodicities related to the solar cycle, apart from the well known synodic 
rotation period of $\sim$27 d and the sunspot activity cycle of $\sim$11 yr, 
have been a topic of interest for over two decades now.  Using data from the Gamma Ray 
Spectrometer \citep{FoC80} onboard the Solar Maximum Mission, a clear 154 d periodicity in 
the occurrence rate of high energy (0.3-100 MeV) flares has been reported \citep{RiK84}.  
This 154 d periodicity in the solar flare occurrence was also confirmed at other 
wavelengths \citep{VJo87,DrG90,KCl91,CRa06}, and for solar energetic particle events 
\citep{Lea90,KSo02,Kil08,CKR10,CDw11}.

In addition, other intermediate periodicities with periods ($<$1 yr) have also been reported 
{\it{viz}}. a 51 d periodicity in the comprehensive flare index \citep{Bai87}, a 127 d periodicity 
in the 10 cm radio flux \citep{KCl91}, periodicities of 33.5 d, 51 d, 84 d, 129 d and 153 d in the 
solar flare occurrence during different phases of solar cycles 19$-$23  \citep{Bai03}, periodicities 
of 100$-$103 d, 124$-$129 d, 151$-$158 d, 177$\pm$2 d, 209$-$222 d, 232$-$249 d, 282$\pm$4 d, 307$-$349 d in 
the unsigned photospheric fluxes \citep{KSB05} and periodicities of 87-106 d, 159-175 d, 194-219 d, 
292$-$318 d, and 69$-$95 d, 113$-$133 d, 160$-$187 d, 245$-$321 d, 348$-$406 d in sunspot areas at different 
phases of cycle 22 and 23 respectively \citep{CKR09}.
 
For periodicities, between 1$-$5 yr, a 1.3 yr periodicity in sunspot areas and numbers 
\citep{KSo02}, 1 yr and 1.7 yr periodicity in solar total and open fluxes \citep{MVV06}, 1.3 yr, 
1.43 yr, 1.5 yr, 1.8 yr, 2.4 yr, 2.6 yr, 3.5 yr, 3.6 yr and 4.1 yr in unsigned photospheric 
fluxes \citep{KSB04,KSB05} have been reported. Studies of all these periodicities have shown the 
existence of an asymmetry in solar activity phenomena between the northern and southern hemispheres 
of the sun.  This hemispheric asymmetry, on different time scales and at different phases of different 
solar cycles, have been observed in various types of solar activity \citep{How74,Ver87,OBa94,KSB05} 
and have been providing invaluable information about basic underlying physical processes on the sun.  
Since processes like differential rotation of the sun and emergence of magnetic flux determine the 
strength and distribution of solar magnetic activity, investigation of solar cycle related periodic 
or quasi-periodic variations is crucial in understanding the behavior and nature of solar magnetic 
fields.

\section*{Decline in mid to high latitude solar fields}
\label{S-mag-field}

\begin{figure}[p]
\centering
\includegraphics[width=16.0cm,height=16.0cm]{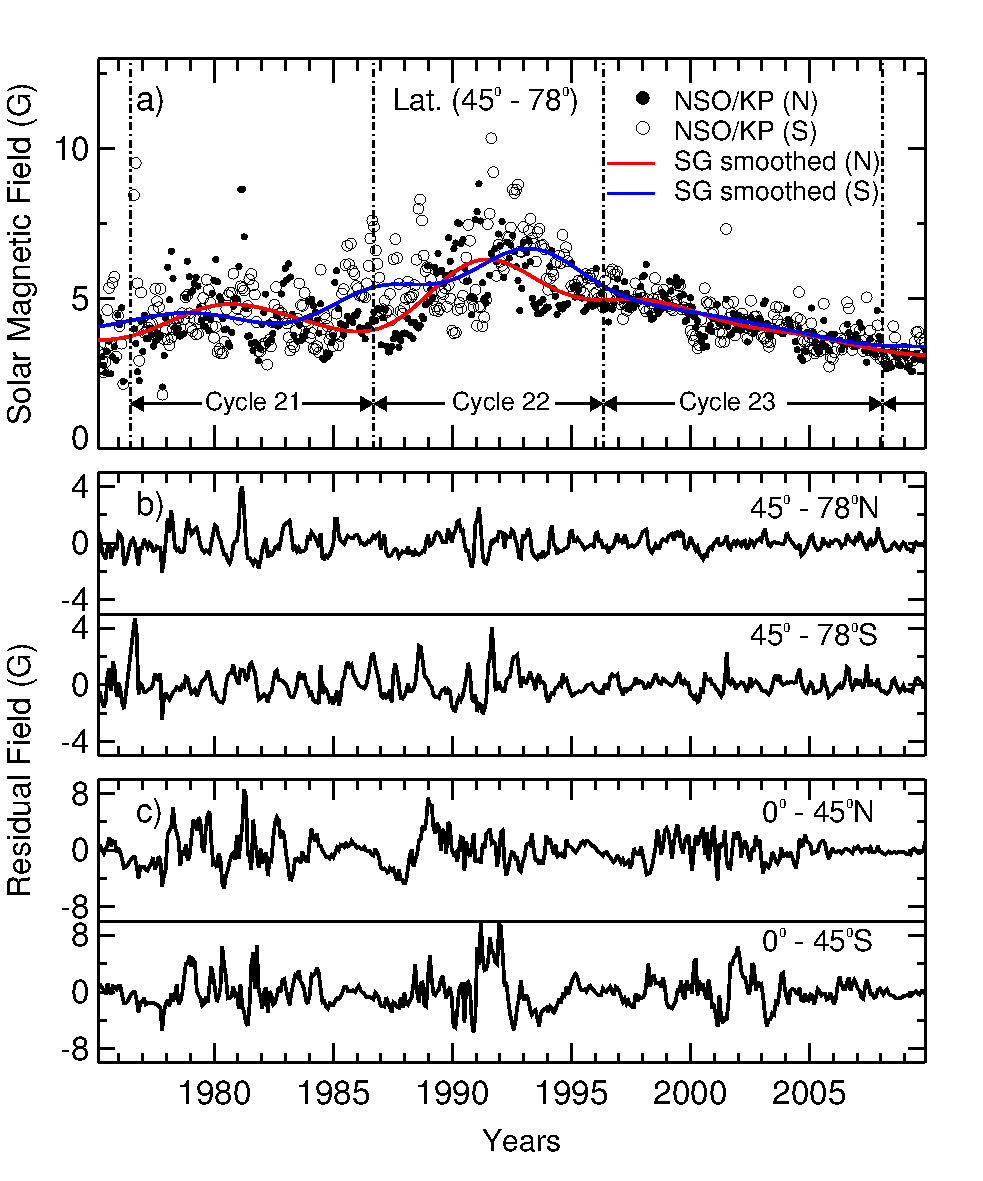}
\caption{ a) Shows measurements of solar photospheric magnetic fields in the latitude range, 
45${^{\circ}}$ to 78${^{\circ}}$.  The filled dots and the open circles respectively represent 
the actual measurements for the northern and southern hemispheres while the fit obtained 
using the SG algorithm is shown by the solid red (north) and blue (south) lines.  The vertically 
oriented dotted lines demarcate the periods of solar cycles 21, 22 and 23.  b) The residuals 
obtained after subtracting the SG smoothed values from the magnetic fields in the latitude range, 
45${^{\circ}}$ to 78${^{\circ}}$, are shown for the north and south.  c) Show the residual fields 
for both the northern and southern hemispheres obtained from the magnetic fields at latitudes 
below 45${^{\circ}}$.}
\label{F-high-lat} 
\end{figure}

Figure \ref{F-high-lat} shows temporal variation in the unsigned or absolute solar photospheric 
magnetic fields in the latitude range, 45${^{\circ}}$$-$78${^{\circ}}$ in both solar hemispheres, 
for the period from February 1975 to November 2009 (panel a) covering solar cycles 21, 22, and 
23.  The vertically oriented dotted lines demarcate the period of solar minimum of these three 
solar cycles.  The small, black filled dots and the open circles are the actual measurements for 
the northern and southern hemispheres respectively while the solid red (northern hemisphere) and 
blue (southern hemisphere) lines are smoothed curves derived using a robust Savitzky-Golay (SG) 
algorithm \citep{SGo64} that has the ability to preserve features of the input distribution like 
maxima and minima while effectively suppressing noise.  Such features are generally flattened by 
other smoothing techniques. The decline in photospheric magnetic fields in both hemispheres, 
starting around the early to mid-1990's can be clearly seen in Fig. \ref{F-high-lat}.
While the decline in photospheric fields had started in $\sim$1991 for the northern hemisphere (solid 
red curve), for the southern hemisphere the decline started in $\sim$1995 (solid blue curve).  
Beyond $\sim$1996, a steady decline in the photospheric fields for both hemispheres can be clearly 
seen.  A similar decline in the sunspot umbral magnetic field strength since $\sim$1998 has also 
been reported \citep{Liv02,PLi06}. Also shown in the bottom two panels of Fig. \ref{F-high-lat} 
are the magnetic residuals, obtained by subtracting the SG fit from the actual measurements, for 
both the hemispheres in the latitude range 45${^{\circ}}$$-$78${^{\circ}}$ (panel b) and 
0${^{\circ}}$$-$45${^{\circ}}$ (panel c) respectively.  See the upper two panels of Fig. 1 in 
\citet{JBG10} for measurements of the magnetic field in the range $\pm$45${^{\circ}}$ which 
show a strong solar cycle modulation in the data as expected.  
%----------------------------  begin Figure 1  -------------------------------------
%\protect\begin{figure}[ht]
%\vspace{14.0cm}
%\special{psfile=Bisoi-etal-f01.eps
%angle=0 hscale=47 vscale=47 voffset=0 hoffset=50}
%-----------------------------end Figure 1------------------------------------------ 
 
Brief details of the method by which the magnetic fields were derived are given below.  We have 
used measurements of magnetic field strengths obtained from 466 individual Carrington rotations 
(CR) of NSO/KP synoptic magnetograms from CR1625 through CR2090 in the period from 19 
February 1975 to 09 November 2009 (1975.14$-$2009.86).  Each magnetogram, generated from daily 
ground-based full-disk magnetograms spanning over a Carrington rotation period corresponding to 
27.27 days, was longitudinally averaged to a 1${^{\circ}}$ wide longitudinal strip covering the 
latitude range from -90${^{\circ}}$ to +90${^{\circ}}$. Surface magnetic fields in both equatorial 
(-45${^{\circ}}$ to +45${^{\circ}}$) and high-latitude zones ($>$45${^{\circ}}$ in both hemispheres) 
of the sun were then derived by averaging over appropriate latitude regions.  Full details of the 
method used are described in \citet{JBG10}.  

The SG smoothing is done in such a way as to make the resulting residual time series stationary 
\textit{i.e.} the mean of time series is or approaches zero.  Such a detrending of the data removes 
large periodic variations ($\sim$11 years).  The residuals were then subjected to both a wavelet 
analysis and a Fourier time series analysis to study temporal periodic variations in photospheric 
magnetic fields.  In the rest of the text the fields obtained in the equatorial belt of -45${^{\circ}}$ 
to +45${^{\circ}}$ are referred to as low-latitude fields while the fields obtained at latitudes 
$>$45${^{\circ}}$ in both hemispheres are referred to as high-latitude fields. 

\subsection*{Transition in Wavelet Periodicities}
  \label{S-wavelet-periods}

A wavelet transform is basically the convolution of a time series with the scaled and 
translated version of a chosen ``${\textit{mother}}$" wavelet function.  The wavelet 
analysis now finds frequent use in the analysis of time series data since it yields 
information in both time and frequency domains \citep{TCo98}.  Using a Morlet wavelet as 
the mother wavelet and based on the algorithm by \citet{TCo98}, the magnetic residuals 
obtained for both the high-latitude and low-latitude fields in the period from 19 February 1975 
to 09 November 2009 (1975.14$-$2009.86) were subjected to a wavelet analysis.  Scattered data 
gaps amounting to $\sim$ 2.5\% of the data were replaced with values obtained using a cubic 
spline interpolation. The Morlet wavelet function $\psi_{0}(\eta)$, represented in equation 
\ref{eq1} below, is a plane wave modulated by a Gaussian.   
\begin{equation}
 \mathsf{ \psi_{0}(\eta) = \pi^{-1/4}e^{i\omega_{0}\eta}e^{-\eta^{2}/2}}
\label{eq1}
\end{equation}
Here, $\omega_{0}$ is a non-dimensional frequency that determines the number of oscillations 
in the wavelet and $\eta$ is a non-dimensional time parameter. The choice of different 
non-dimensional frequencies defines the frequency resolution.  However, by varying the frequency 
resolution, one has to compromise with time resolution. In this case, we have chosen $\omega_{0}$ 
to be 6 for obtaining a better frequency resolution.  This value is left unchanged and is not 
tuned to different values as our interest is to look for changes in the frequency components 
rather than to find precisely the low and high frequency components.  However, in the following sections, 
we have employed a Fourier analysis technique to find the various frequency components. 

\begin{figure}[p]
\centering
\includegraphics[width=12.0cm,height=12.0cm]{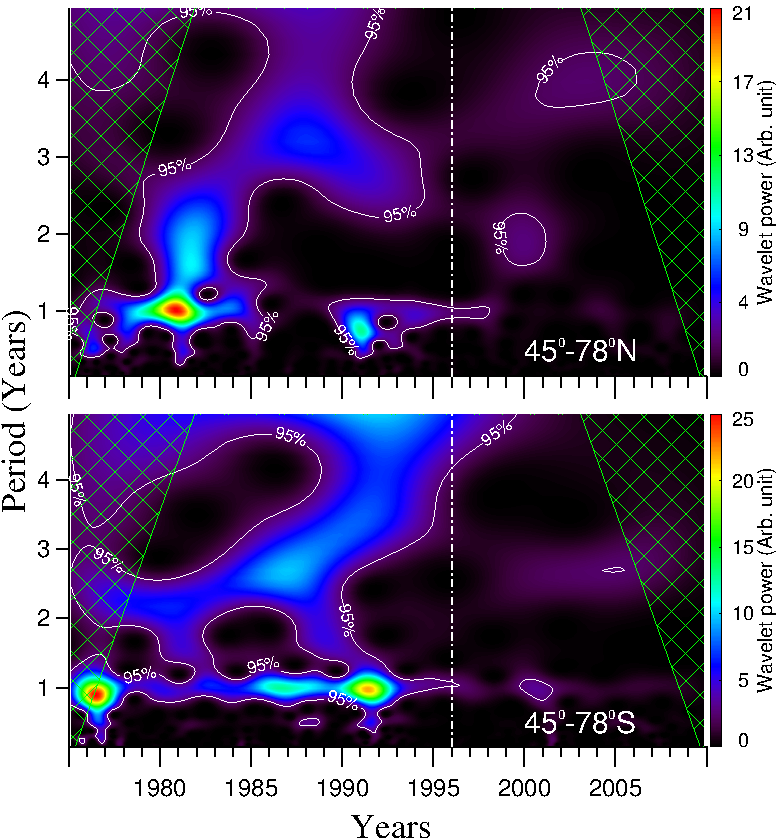}
\caption{Shows the wavelet power distribution of solar periodicity over years for 
high-latitude fields for the northern hemisphere (upper panel) and southern 
hemisphere (lower panel) in the latitude range, 45${^{\circ}}$ to 78${^{\circ}}$.  
The green cross hatched regions are the cone of influence (COI) while the white contours 
indicate a significance level at 95\%.  The dashed vertical lines are marked at 1996 
when a clear transition has observed in wavelet power of periodicities for both the 
north and south high-latitude fields.}
\label{F-wavelet-plots} 
\end{figure}

\begin{figure}[p]
\centering
\includegraphics[width=12.0cm,height=12.0cm]{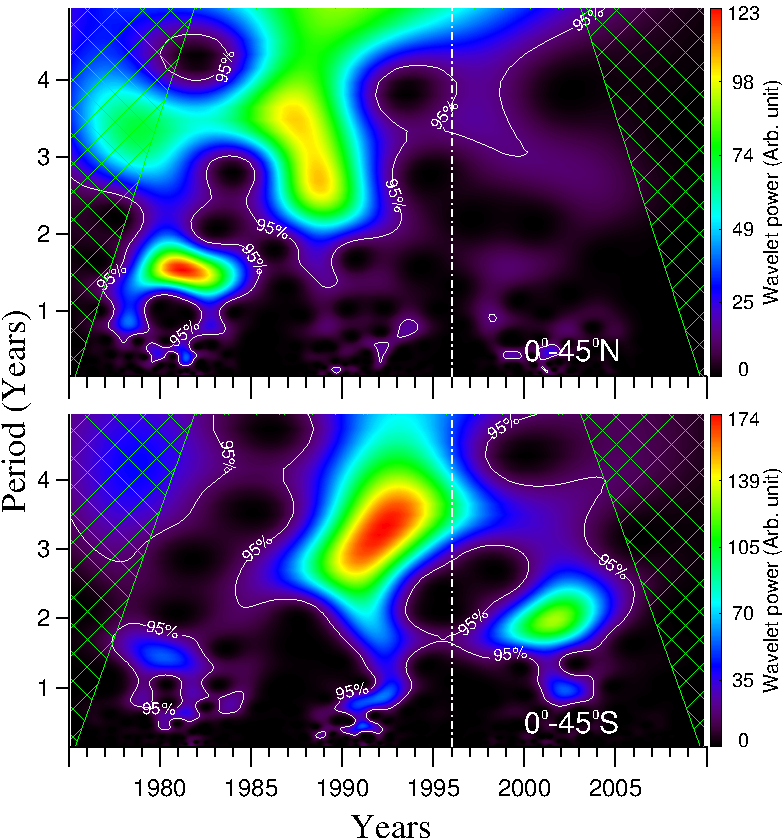}
\caption{Shows the wavelet power distribution of solar periodicity over years for 
low-latitude fields for the northern hemisphere (upper panel) and southern 
hemisphere (lower panel) in the latitude range, 0${^{\circ}}$ to 45${^{\circ}}$. 
The green cross hatched regions are the cone of influence (COI) while the white 
contours indicate a significance level at 95\%.  The dashed vertical line demarcates 
the period of transition in wavelet power of periodicities for both the hemispheres 
around 1996.}
\label{F-wavelet-plots-tor} 
\end{figure}

Figure \ref{F-wavelet-plots} shows the wavelet spectrum for the high-latitude fields in the 
northern (upper panel) and southern hemisphere (lower panel) while Figure \ref{F-wavelet-plots-tor} 
shows the wavelet spectrum for the low-latitude fields in the northern (upper panel) and 
southern hemispheres (lower panel).  The green cross-hatched regions in both figures represent 
the cone of influence (COI) where the spectral distribution is not reliable. The white contours 
are drawn at a significance level of 95\%.  The power levels are indicated by a color coded bar 
on the right of each panel and the dashed vertical line in each panel is drawn at 1996 in both 
Figures \ref{F-wavelet-plots} and \ref{F-wavelet-plots-tor}.  For the high-latitude fields 
(Fig. \ref{F-wavelet-plots}), in both the north and south, significant wavelet power is only 
seen at lower periods ($\sim$1$-$3 yr) prior to 1985 while significant wavelet power is seen 
both at lower and higher periods ($\sim$3$-$5 yr) for the interval between 1985 and 1996.  For the 
low-latitude fields (Fig. \ref{F-wavelet-plots-tor}), we don't see significant changes in the 
power levels between 1975 and 1996. 
%
%-------------------------begin Figure 2------------------------------------------
%\protect\begin{figure}[ht]
%\vspace{11.2cm}
%\special{psfile=Bisoi-etal-f02.eps
%angle=0 hscale=50 vscale=50 voffset=0 hoffset=80}
%-------------------------------end Figure 2-------------------------------------

It is important to note here that there is a clear transition in both periodicities and power 
levels in both hemispheres around 1996 for both high-latitude and low-latitude fields.  We 
stress here that we have taken care to check that the selection of different colors or changes 
in the color intensity in the wavelet plots makes no difference in the transition noticed in 
the wavelet power spectra.  Though the decline in the fields has started in the early to mid 
1990's, as seen from Fig. \ref{F-high-lat}, we see from the wavelet spectra that there is an 
unambiguous transition in the quasi-periodic variations of fields in both the hemispheres at 
$\sim$1996.  We have therefore chosen, in the rest of the analysis, to divide the residuals 
into two sets based on this transition seen in the wavelet spectra around $\sim$1996 in order 
to study the changes, if any, in the periodicities before and after the transition seen in 
the wavelet spectra. A Fourier time series analysis was then carried out separately on the 
residuals used in the wavelet analysis for the period prior to 1996 and the period after 1996 
for both high-latitude and low-latitude fields. 

It is important to bear in mind here that the transition seen in the wavelet 
spectra around 1995$-$1996 corresponds to the time when both solar high-latitude fields 
above $\pm$45${^{\circ}}$ and solar wind turbulence levels in the entire inner-heliosphere 
began declining \citep{JBG10,JaB11}.
%
%------------------------------------begin Figure 3----------------------------
%\protect\begin{figure}[ht]
%\vspace{10.0cm}
%\special{psfile=Bisoi-etal-f03.eps
%angle=0 hscale=50 vscale=50 voffset=0 hoffset=100}
%----------------------------------end Figure 3-------------------------------------
%  
\section*{Fourier periodicities - Asymmetries and symmetries}
   \label{S-fourier-periods}   
   
Based on the transition seen in the wavelet spectra at $\sim$1996, we have subjected 
the residuals obtained from the SG fits to a Fourier analysis after dividing the time 
series of the residuals, spanning years 1975.14 to 2009.86, into two parts, one prior to 
1996 and the other after 1996.  Since the time series of the residual field has some data 
gaps amounting to $\sim$2.5\% of the data, the algorithm used for this purpose was based 
on the Lomb-Scargle Fourier Transform for unevenly spaced data \citep{Lom76,Sca82,Sca89,SSt97} 
in combination with the Welch-Overlapped-Segment-Averaging (WOSA) procedure \citep{Wel67}.  
The spectral power derived from such an analysis is generally normalized with respect to 
the total power contained in all the Fourier components taken together.  This procedure makes 
the relative distribution of the spectral power independent of the spectral windowing used 
in the algorithm. Such normalized power spectra are referred to as normalized Fourier 
periodograms or simply normalized periodograms.   
  
\subsection*{High-latitude Fields}
\label{S-asym-high-latitude}        

Normalized periodograms obtained from the Fourier analysis are shown in Figure \ref{F-asym-fields}.  
The top four panels of Fig. \ref{F-asym-fields} labeled a, b, c and d represent the 
high-latitude fields while the bottom four panels of Fig. \ref{F-asym-fields} labeled 
e, f, g, and h represent the low-latitude fields.  The upper four panels (high-latitude 
fields) show normalized periodograms before and after 1996 in the north (panels a and b 
respectively), and before and after 1996 in the south (panels c and d respectively).  In 
a similar fashion, the lower four panels (low-latitude fields) show normalized periodograms 
in the north before and after 1996 (panels e and f respectively) and in the south before 
and after 1996 (panels g and h respectively).  The ``${significant}$" periodic components 
were determined using the Siegel test statistics \citep{Sie80} and are those components 
having power levels above the black horizontal line drawn in each panel of Fig. \ref{F-asym-fields}. 
The Siegel's test is an extension of the Fisher's test \citep{Fis29}. While Fisher's test 
attempts to spot the single dominant periodicity in a time series that has maximum power 
in the periodogram and is above the ``${critical}$" level defined by Fisher's test statistics, 
Siegel's test \citep{Sie80} relaxes this stringent condition a little and considers two to 
three dominant periodic components above the ``${critical}$" level. Both the 
aforementioned test statistics and the process of determining the confidence/significance 
level are discussed in \citep{PWa93} and also briefly in \citep{SSt97}. In the present analysis, 
95\% confidence level (or 5\% significance level) is chosen. This is equivalent to $\pm$2$\sigma$ 
level in FFT method. It is to be noted that the choice of this confidence level decides the 
``${critical}$" level of the Sigel test statistics \citep{SSt97}. 

It may be noted that though the Fourier components obtained from the analysis 
range in periods from 54 d to 12 yr corresponding to 214 nHz to 2.6 nHz in 
frequency, Fig. \ref{F-asym-fields} only shows those components with periods lying 
in the range 300 d to 12 yr ${\it{i.e.}}$ the low-frequency range.  Normalized 
periodograms containing Fourier components with periods less than 300 d will be 
discussed later. 

The red vertical lines in each panel of Fig. \ref{F-asym-fields} are drawn through 
the peak of the component with the longest period.  A shift in the periodicities 
can be seen after 1996 when compared to that before 1996.  This shift is indicated 
by vertical dashed blue lines and a black arrow showing the direction of the shift 
in periodicities.  It must be emphasized that we are not interested in the actual 
periodicities themselves but rather in the manner in which the longest periodicities 
shift between the time interval prior to and after 1996. It can be seen from 
Fig. \ref{F-asym-fields} that in addition to the shift in periodicity of the high-latitude 
fields in both hemispheres before and after 1996 the spectral power also differs 
significantly before and after 1996. The longer periods after 1996 have more spectral 
power than longer periods prior to 1996. If we denote the longest periods for the 
high-latitude fields in the north before 1996 as $T_1$ and that after 1996 as $T_2$, 
we find that $T_2$ $>$ $T_1$ for high-latitude fields in the north. 

For the southern hemisphere, the direction of the shift in periodicities, as indicated 
by the direction of the black arrow in Fig. \ref{F-asym-fields}, is in the opposite 
direction as compared to the northern hemisphere.  If we denote the longest periods for 
the high-latitude fields in the south before 1996 and after 1996 as $T_3$ and $T_4$ 
respectively, we find that $T_4$ $<$ $T_3$ for the high-latitude fields in the south.  
As in the case of the northern hemisphere, the distribution of spectral power in various 
periodicities is also significantly different in the periods before and after 1996.  

Thus, a north-south asymmetry is seen in the distribution of periodicities for the 
high-latitude fields along with significant changes in the power of various periodic 
components.  Since the normalized periodograms are generated from measurements of 
magnetic fields, it is a proxy for the behaviour of the photospheric magnetic fields 
and a north-south asymmetry would imply a similar asymmetry in photospheric magnetic 
fields towards higher solar latitudes before and after 1996.

\subsection*{Low-latitude fields}
             \label{S-asym-torr}
             
\begin{figure}[p]
\centering
\includegraphics[width=14.0cm,height=16.0cm]{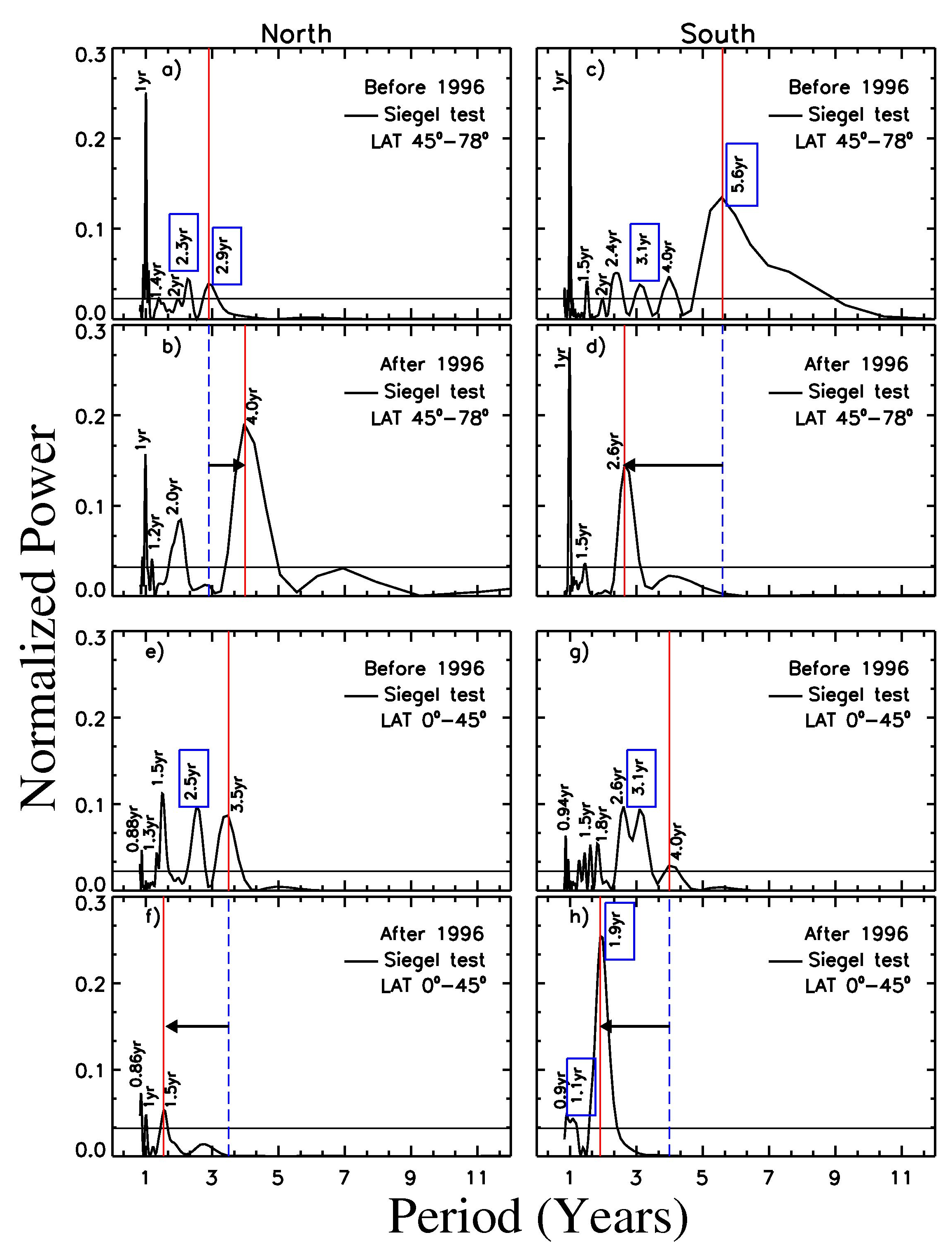}
\caption{The top four panels - a, b, c, d and the bottom four panels - e, f, g, h 
respectively represent normalized Fourier periodograms for high-latitude fields 
and low-latitude fields.  The four panels in the top and the bottom show respectively, 
the normalized Fourier power distribution with periodicity for photospheric fields 
in the north before 1996, in the north after 1996, in the south before 1996 and 
in the south after 1996.  The solid red vertical lines demarcate the highest period 
in each of the four panels while the dotted blue vertical line and direction of the 
black arrow in each panel are used to show the shift in the periodicity of the longest 
periods.  The solid horizontal lines depict significance levels as determined by Siegel 
test. The new periodicities from our analysis have been boxed in blue.}
\label{F-asym-fields}
\end{figure}

In contrast to the behavior of the high-latitude fields, the low-latitude fields 
represented in the bottom four panels of Fig. \ref{F-asym-fields} and labeled e, f, g and h 
show no asymmetry in the manner in which the longest periods shift when one compares 
the periodograms prior to and after 1996.  The shift in the longest periods in both the 
north and south low-latitude fields before and after 1996 has been depicted in the 
lower four panels of Fig. \ref{F-asym-fields} in a similar fashion to that of the upper 
four panels.  The vertical solid red line is drawn through the peak of the component 
with the longest period and the dotted blue line and a black arrow in each panel 
indicate the direction of the shift.  

Denoting the longest period for the low-latitude fields in the north before and after 1996 
as $T_5$ and $T_6$ respectively, we find that $T_5$ $>$ $T_6$.  In the southern hemisphere, 
denoting the longest period for the low-latitude fields prior to 1996 and after 1996 as $T_7$ and 
$T_8$ respectively, it is clear that $T_7$ $>$ $T_8$.  Hence, the longest periods show a shift in 
the same direction indicating no asymmetry in the distribution of the low-latitude fields before 
and after 1996.
%
%%------------------------------------begin Figure 4---------------------------------------
%\protect\begin{figure}[ht]
%\vspace{12.8cm}
%\center
%\special{psfile=Bisoi-etal-f04-new.eps
%angle=0 hscale=38 vscale=30 voffset=0 hoffset=60}
%%----------------------------------end Figure 4-----------------------------------
%%

\section*{Latitudinal profile for Fourier power spectrum}
       \label{S-lat-profile}
       
To examine how the Fourier spectrum, for the high-latitude fields, varies as a function 
of latitude, the residuals for the high-latitude fields were further subdivided into 
15${^{\circ}}$ wide latitude bins and a Fourier analysis was carried out for the 
data in each latitude bin.  This was done to examine the changes, if any, in the spectral 
distribution with latitude and because the strength and distribution of photospheric 
magnetic fields change with solar latitudes.  It would therefore be interesting 
to see if the north-south asymmetry in the high-latitude fields persists across all 
latitude bins.  Figure \ref{F-prof-high} shows normalized periodograms in the
\noindent latitude range 30${^{\circ}}$ to 75${^{\circ}}$ in latitude steps of 
15${^{\circ}}$.  The top four panels of Fig. \ref{F-prof-high} labeled a, b, c, d, the 
middle four panels labeled e, f, g, h, and the bottom four panels labeled i, j, k, l 
represent respectively, normalized periodograms showing the distribution of Fourier 
periodicities in the latitude ranges 30${^{\circ}}$$-$45${^{\circ}}$, 45${^{\circ}}$$-$60${^{\circ}}$ 
and 60${^{\circ}}$$-$75${^{\circ}}$.  The latitude ranges are indicated at the top right 
hand corner of each panel.  Starting from the top, each of the left hand panels of 
Fig.\ref{F-prof-high} represent, in pairs before 1996 and after 1996 respectively, the power 
spectra for the northern hemisphere in each latitude bin. Similarly, the right hand panels 
of Fig.\ref{F-prof-high} represent, in pairs before 1996 and after 1996 respectively, the power 
spectra obtained for each latitude bin in the southern hemisphere. In a manner similar to 
that shown in Fig. \ref{F-asym-fields}, the red vertical lines in each panel of Fig. \ref{F-prof-high} 
are drawn through the peak of the component with the longest period and the dotted blue line 
and a black arrow indicate the direction in which the longest periodicities shift.  It can 
be seen from Fig.\ref{F-prof-high} that the north-south asymmetry is present only in the 
45${^{\circ}}$$-$60${^{\circ}}$bin.  Though not shown in Fig. \ref{F-prof-high} we have 
verified that the asymmetry is also not there in the latitude bins 0${^{\circ}}$ to 
15${^{\circ}}$ and 15${^{\circ}}$ to 30${^{\circ}}$.

In the 60${^{\circ}}$$-$75${^{\circ}}$ bin (bottom four panels of the Fig \ref{F-prof-high}), 
the shift of the longest periods before and after 1996 shows no north-south asymmetry. 
However, the spectral powers in the longest periods are significantly different before 
and after 1996.  The spectral power in the longest period component increases significantly 
after 1996 for the northern hemisphere while it decreases for the southern hemisphere. 

It is important to note here that the latitude band 45${^{\circ}}$$-$60${^{\circ}}$ in both 
hemispheres is dominated by surges or tongues of magnetic flux being carried polewards by 
the meridional flow.  These surges are a direct surface manifestation of the meridional 
flow which is in turn governed by an internal solar dynamo.  Beyond 60${^{\circ}}$ in 
latitude, the high-latitude flux in both hemispheres saturates and can be best seen along 
with the magnetic surges in the 45${^{\circ}}$$-$60${^{\circ}}$ latitude band in a magnetic 
butterfly diagram which will be described in the following section.

\subsection*{The Magnetic butterfly diagram and polar surges}

It is known that the emergence and evolution of magnetic field on the solar photosphere 
is tied in to a solar dynamo operating in the solar interior. In particular, the toroidal 
field manifests itself as solar surface bipolar active regions with sunspots migrating toward 
the equator as the solar cycle progresses.  It is this migration which gives rise to the well 
known ``{\textit{butterfly}}" diagram which is basically a map of longitudinally averaged magnetic 
fields in time.  Solar dynamo models that explain the main features of solar magnetic activity 
consider the strong toroidal field as being generated by the shearing of a pre-existing weak poloidal 
field by solar differential rotation ($\omega$ effect).  Subsequently, the regeneration and 
inversion of polar fields, at the maximum of each solar cycle, is caused by the cancellation 
of sunspot fields at the equator along with a net poleward surface flux that is transported 
via meridional circulation to reach the poles and reverse the pre-existing polar fields
(Babcock-Leighton type $\alpha$-effect).
%%------------------------------------begin Figure 5----------------------------------
%\protect\begin{figure}[p]
%\vspace{12.0cm}
%\center
%\special{psfile=Bisoi-etal-f05-new.eps
%angle=0 hscale=35 vscale=24 voffset=0 hoffset=70}
\begin{figure}[p]
\centering
\includegraphics[width=14.0cm,height=16.0cm]{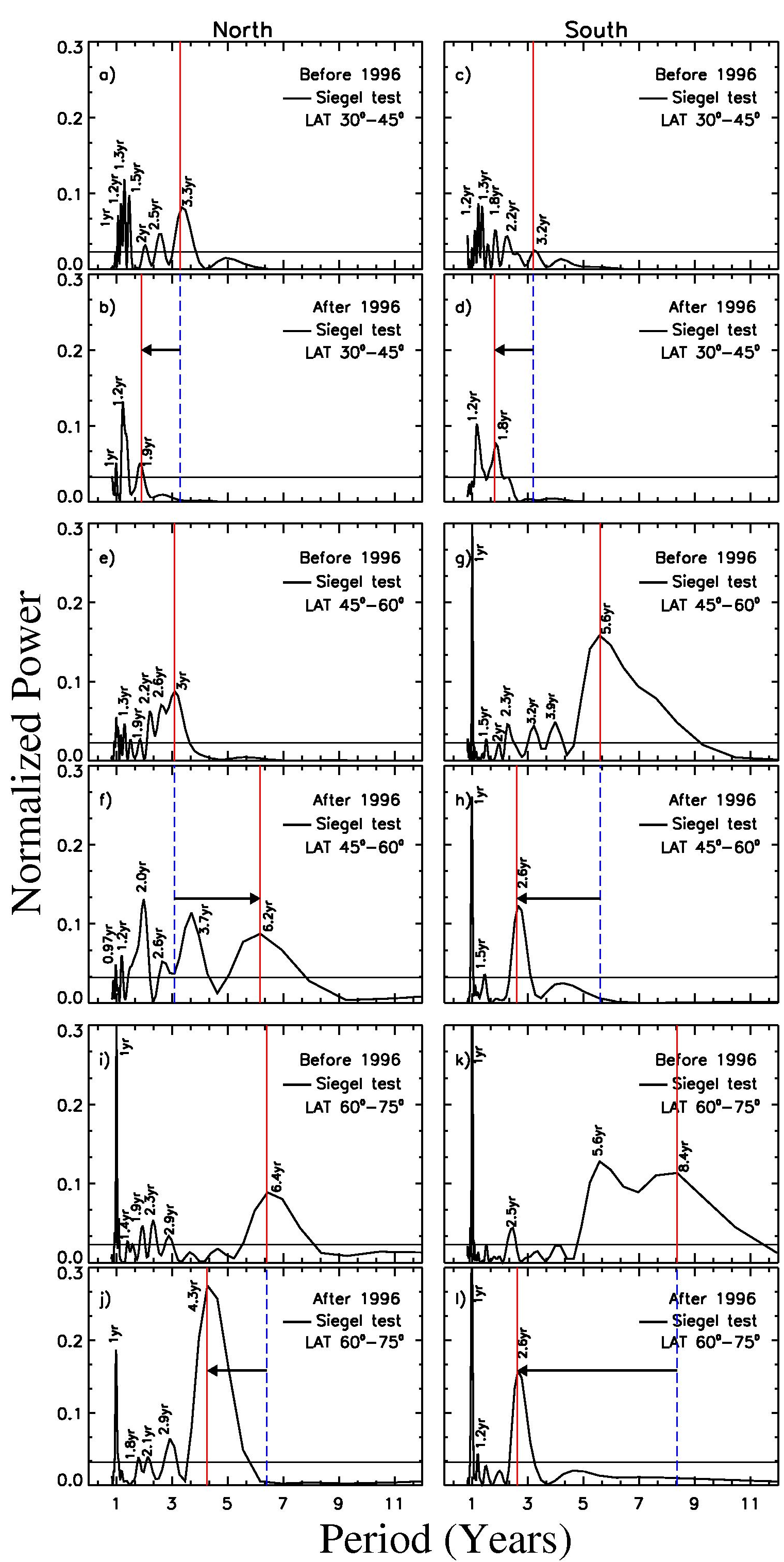}
\caption{ The top four panels - a, b, c, d, the mid - e, f, g, h, and the bottom 
four panels - i, j, k, l respectively represent the normalized Fourier periodograms 
for the photospheric fields in the latitude ranges, 30${^{\circ}}$$-$45${^{\circ}}$,  
45${^{\circ}}$$-$60${^{\circ}}$, and 60${^{\circ}}$$-$75${^{\circ}}$.  The four panels 
(the top, mid and bottom) in the figure correspond to the Fourier spectral power for 
magnetic fields in the north before 1996, for magnetic fields in the north after 1996, 
for magnetic fields in the south before 1996, and for magnetic field in the south after 
1996.  The solid red vertical line, the dotted blue vertical line and the black arrow in 
each panel are used to show the shift in periodicity and the spectral power of the fields 
before and after 1996 both in the northern and southern hemisphere.  The solid horizontal 
lines depict significance levels as determined by Siegel test.}
\label{F-prof-high}
\end{figure}
%%----------------------------------end Figure 5----------------------------------------------
%  

We have generated a butterfly diagram from the NSO/Kitt-Peak synoptic maps by using 
longitudinally averaged photospheric fields derived from the synoptic maps described in 
section \ref{S-mag-field}.  Figure \ref{F-butterfly} shows a butterfly diagram for solar cycles 
21, 22 and 23 covering the period from 19 February 1975 to 09 November 2009 
(1975.14~$-$~2009.86).  The leading and following polarity fluxes in panel - a of Fig. 
\ref{F-butterfly} are shown in red and green respectively for the northern hemisphere in cycle 21, 
with the derived flux values indicated in a colour coded bar at the top of the figure.
It must be noted that leading and following polarity fluxes will be reversed in the southern 
hemisphere and will again flip from cycle to cycle and hemisphere to hemisphere. 
 
In order to obtain better contrast, the fluxes in Fig. \ref{F-butterfly}a have been limited 
to the range $\pm30$G. The poleward motion of the trailing polarity fluxes, above the 
latitudes ${\pm}$45${^{\circ}}$, known as polar surges are episodic in nature \citep{WNS89} and 
can be clearly seen as tongues of red and green bands in the upper panel of Fig. \ref{F-butterfly} 
labeled - a.  For a better view of the polar surges, the middle and lower panels of Fig. 
\ref{F-butterfly} show variations of intensity as a function of time in each 1${^{\circ}}$ strip 
of latitude in the latitude range 45${^{\circ}}$$-$60${^{\circ}}$ for the northern hemisphere 
(middle panel - b) and southern hemisphere (lower panel - c).  The solar field for each degree 
of latitude has been shown after smoothing over three Carrington rotations. An example of a 
surge in the north and south has been highlighted in red.  

\cite{KSB05} reported a periodicity of 1.3 yr in the photospheric fields that has 
been correlated with these polar surge motions.  We see an approximate periodicity of 
1 yr in the polar surge flows in the lower two panels of Fig. \ref{F-butterfly}.  This agrees 
with the strong 1 yr periodicity seen in Fig. \ref{F-asym-fields} and \ref{F-prof-high}.  Further, 
a variation in the strength and occurrence rate of surges prior to and after 1996 can be seen 
from a careful inspection of Fig. \ref{F-butterfly}a, b and c.  In the latitude bin 
${\pm}$45${^{\circ}}$ $-$ ${\pm}$60${^{\circ}}$, we find that the frequency of these surges 
are comparatively more during the years 1996$-$2009 while it is less during the years 1986$-$1995.

\section*{Fourier periodicity - Low and high frequency periods}
   \label{S-fourier-period}
The harmonic analysis yielded Fourier components with periods ranging from 54 d to 
12 yr corresponding to a frequency range of 2.6 nHz to 214 nHz.  The significant 
periodic components, as determined by Siegel test, were grouped into low frequency 
periods in the frequency range, 38.5 nHz (300 d) to 2.6 nHz (12 yr) and high 
frequency periods in the frequency range, 214 nHz (54 d) to 38.5 nHz (300 d).  
These periods have been listed in Table \ref{T-tab-low} (long periods) and Table 
\ref{T-tab-high} (short periods) respectively.  Many of these periodicities 
have already been reported by other workers and a few new periodicities have also 
been seen in our analysis.  These new periodicities have been listed in bold face in 
Tables \ref{T-tab-low} and \ref{T-tab-high} and boxed in blue in Figures \ref{F-asym-fields} 
and \ref{F-asym-fields-high}.

\subsection*{The low frequency periods}
     \label{S-period-low} 
In the low frequency range, 38.5 nHz (300 d) to 2.6 nHz (12 yr), we restrict 
our analysis to periodicities in the range, 305 d to 5.6 yr.  Periodicities greater 
than 5.6 yr are not taken into account because of their inaccurate determination as a 
consequence of division of time series.  The periodicities are listed in Table \ref{T-tab-low} 
where the upper half lists periodicities for high-latitude fields in the north and south 
grouped before and after 1996 while the lower half lists periodicities for low-latitude 
fields in a similar fashion. 

For photospheric fields in the high-latitude range we find periodicities of 305 d (0.84 yr), 
325 d (0.89 yr), 339 d (0.92yr), 341 d (0.93 yr), 1 yr, 1.2 yr, 1.4 yr, 1.5 yr, 2 yr, 2.4 yr, 
2.6 yr, and 4 yr while for photospheric fields in the low-latitude range we find periods of 
312 d (0.86 yr), 318 d (0.87 yr), 321 d (0.88 yr), 327 d (0.9 yr), and 343 d (0.94 yr), 1 yr, 1.27 yr,
1.3 yr, 1.44 yr, 1.5 yr, 1.8 yr, 2.6 yr, 3.5 yr, and 4 yr.  All of these quasi-periodicities in 
various types of solar activity have been reported earlier by other researchers 
\citep{HoC00,KSo02,KSB05,MVV06,CKR09}.  \cite{KSB05} had discussed a north-south asymmetry in 
quasi-periodic variations in unsigned photospheric flux with periods of 1.3yr, 1.5 yr, and 2.6 yr. 
%
%------------------------------------begin Figure 6-------------------------------------
%\protect\begin{figure}[ht]
%\vspace{12.0cm}
%\center
%\special{psfile=Bisoi-etal-f06.eps
%angle=0 hscale=35 vscale=21 voffset=0 hoffset=70}
\begin{figure}[p]
\centering
\includegraphics[width=12.0cm,height=12.0cm]{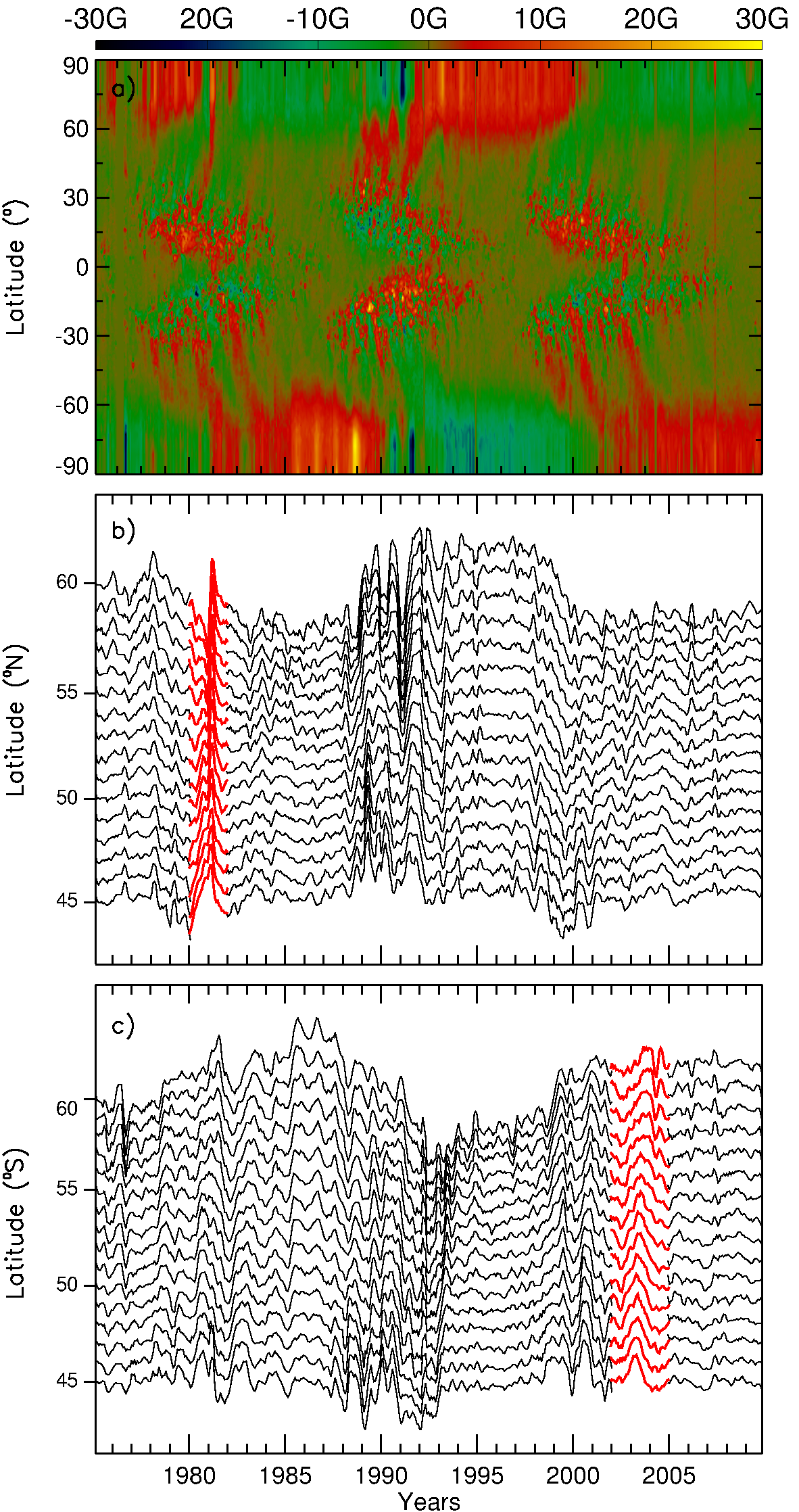}
\caption{The top panel labeled - a shows a magnetic butterfly diagram generated 
using NSO/Kitt-Peak magnetograms and depicts the net photospheric magnetic flux 
distribution on the sun for solar cycles 21, 22 and 23.  For better contrast 
the magnetic flux has been limited to 30 Gauss. The positive and negative polarities 
are shown in red and green respectively.  Polar surges or lateral motions of magnetic 
flux moving polewards above latitudes of $\pm$45${^{\circ}}$ are seen as tongues 
of red and green bands.  The lower two panels labeled - b and c show the variation 
of magnetic field, after smoothing over three Carrington rotations, for each 
degree of latitude in the range 45${^{\circ}}$ to 60${^{\circ}}$.  An example of a 
surge is highlighted in red in the bottom two panels representing the northern and 
southern hemisphere respectively.}
\label{F-butterfly}
\end{figure}
%------------------------------------end Figure 6--------------------------------------------
    
In addition to these known periodicities, we have found some new periodicities in the high-latitude 
fields with periods 2.3 yr, 2.9 yr, 3.1 yr, and 5.6 yr and in low-latitude fields 
with periods 1.1 yr, 1.9 yr, 2.5 yr, and 3.1yr.  These new periodicities have been shown 
in boldface in Tables \ref{T-tab-low} and \ref{T-tab-high}.  Though the periodicity of 1 yr has already been 
reported, we find this periodicity to be the prominent periodicity observed in both north and south 
high-latitude fields before and after 1996.  The Fourier power in the 1 yr period is comparatively more 
both in the north and south prior to 1996 than after 1996. On the other hand, though the 1 yr period is 
present in the low-latitude fields both in the north and south, it is only present after 1996.  Besides
these, periods 1.2$-$1.6 yr is common to both the north and south low-latitude fields and have significant power.  
%------------------------Begin Table 1 1996--------------------------------------
\begin{sidewaystable}
\hspace{2.0cm}
\resizebox{16cm}{5cm}{
\begin{tabular}{ccc|cc|cc|ccc}
\hline
&\multicolumn{4}{c|}{North High-latitude Field} &\multicolumn{4}{c}{South High-latitude Field}\\
\hline
&\multicolumn {2}{c|}{Before 96}  &\multicolumn {2}{|c|}{After 96} &\multicolumn {2}{|c|}{Before 96} 
&\multicolumn {2}{|c}{After 96}\\
\hline
&Period(yr) &Power  &Period(yr) &Power  &Period(yr)  &Power  &Period(yr)  &Power\\
\hline
&$\bf{2.9}$   &$\bf{0.04}$  &4.0  &0.19  &$\bf{5.6}$  &$\bf{0.13}$  &2.6  &0.15\\
&$\bf{2.3}$   &$\bf{0.04}$  &2.0  &0.08  &4.0  &0.05  &1.5  &0.04\\
&2.0   &0.02  &1.2  &0.04  &$\bf{3.1}$  &$\bf{0.04}$  &1.0  &0.28\\
&1.4   &0.02  &1.0  &0.16  &2.4  &0.05  &---  &---\\
&1.0   &0.25  &0.89  &0.04  &2.0  &0.02  &---  &---\\
&0.93  &0.06  &---  &---  &1.5  &0.04  &---  &---\\
&---  &---  &---  &---  &1.0  &0.02  &---  &---\\
&---  &---  &---  &---  &0.92  &0.03  &---  &---\\
&---  &---  &---  &---  &0.89  &0.02  &---  &---\\
&---  &---  &---  &---  &0.84  &0.04  &---  &---\\
&---  &---  &---  &---  &---  &---  &---  &---\\
\hline
&\multicolumn{4}{c|}{North Low-latitude Field} &\multicolumn{4}{c}{South Low-latitude Field}\\
\hline
&\multicolumn {2}{c|}{Before 96}  &\multicolumn {2}{|c|}{After 96} &\multicolumn {2}{|c|}{Before 96} 
&\multicolumn {2}{|c}{After 96}\\
\hline
&Period(yr)  &  Power  &Period(yr) &  Power  &Period(yr)  &  Power  &Period(yr)  &  Power\\
\hline
&3.5  &0.09  &1.5  &0.05  &4.0  &0.03  &$\bf{1.9}$  &$\bf{0.25}$\\
&$\bf{2.5}$  &$\bf{0.10}$  &1.0  &0.04  &$\bf{3.1}$  &$\bf{0.09}$  &$\bf{1.1}$  &$\bf{0.04}$\\
&1.5  &0.11  &0.86  &0.07  &2.6  &0.1  &0.90  &0.05\\
&1.3  &0.04  &---  &---  &1.8  &0.05  &---  &---\\
&0.88  &0.04  &---  &---  &1.5  &0.05  &---  &---\\
&---  &---  &---  &---  &1.44    &0.04  &---  &---\\
&---  &---  &---  &---  &1.27    &0.04  &---  &---\\
&---  &---  &---  &---  &0.94    &0.04  &---  &---\\
&---  &---  &---  &---  &0.87    &0.06  &---  &---\\
\hline
\end{tabular}
}
\caption{ The Fourier periods in the frequency range, 38.5 nHz to 2.6 nHz (Low frequency 
zone), are listed with their Fourier power respectively in the upper and lower half of 
the table for high-latitude fields and low-latitude fields in both the north and south 
prior to 1996 and after 1996.}
\label{T-tab-low}
\end{sidewaystable}
%%----------------------------End Table 1 1996---------------------------------------
%%%
%------------------------------Begin Table 2 1996-------------------------------------
\protect\begin{sidewaystable}
\hspace{2.0cm}
\resizebox{16cm}{5cm}{
\begin{tabular}{ccc|cc|cc|ccc}
\hline
&\multicolumn{4}{c|}{North High-latitude Field} &\multicolumn{4}{c}{South High-latitude Field}\\
\hline
&\multicolumn {2}{c|}{Before 96}  &\multicolumn {2}{|c|}{After 96} &\multicolumn {2}{|c|}{Before 96} 
&\multicolumn {2}{|c}{After 96}\\
\hline
&Period(day) &Power  &Period(day) &Power  &Period(day)  &Power  &Period(day)  &Power\\
\hline
&$\bf{294}$   &$\bf{0.02}$  &$\bf{253}$  &$\bf{0.04}$  &288  &0.037  &184  &0.07\\
&278   &0.06  &183  &0.061  &184  &0.06  &---  &---\\
&263   &0.04  &---  &---  &---  &--- &---  &---\\
&240   &0.02  &---  &---  &---  &---  &---  &---\\
&183   &0.04  &---  &---  &---  &---  &---  &---\\
&157   &0.04  &---  &---  &---  &---  &---  &---\\
\hline
&\multicolumn{4}{c|}{North Low-latitude Field} &\multicolumn{4}{c}{South Low-latitude Field}\\
\hline
&\multicolumn {2}{c|}{Before 96}  &\multicolumn {2}{|c|}{After 96} &\multicolumn {2}{|c|}{Before 96} 
&\multicolumn {2}{|c}{After 96}\\
\hline
&Period(day)  &  Power  &Period(day) &  Power  &Period(day)  &  Power  &Period(day)  &  Power\\
\hline&288  &0.02  &$\bf{257}$  &$\bf{0.04}$  &283  &0.04  &---  &---\\
&278  &0.03  &181  &0.08  &248  &0.02  &---  &---\\
&261  &0.03  &172  &0.11  &228  &0.03  &---  &---\\
&---  &---  &147  &0.04  &157  &0.03  &---  &---\\
&---  &---  &---  &---  &151  &0.03  &---  &---\\
&---  &---  &---  &---  &145  &0.03  &---  &---\\
&---  &---  &---  &---  &---  &---  &---  &---\\
&---  &---  &---  &---  &---  &---  &---  &---\\
\hline
\end{tabular}
}
\caption{The Fourier periods with their Fourier spectral power in the frequency 
range, 214 nHz to 38.5 nHz (High frequency zone) are listed respectively in the 
upper and lower half of the table for high-latitude fields and 
low-latitude fields both in the north and south prior to 1996 and after 1996.}
\label{T-tab-high}
\end{sidewaystable} 
%----------------------------------------End Table 2 1996---------------------------------------

Among these periodicities, the most discussed and reported periodicity is the 1.3 yr periodicity 
\citep{KSo02,KSB04,KSB05} which has been linked to polar surges that can be seen most clearly in the 
latitude band 45${^{\circ}}$ to 60${^{\circ}}$ \citep{WNS89,KSB05} in magnetic butterfly diagrams 
as depicted in Fig. \ref{F-butterfly}.  Also, the 1.3 yr periodicity has been linked to the 
variation of the rotation rate near the base of the convection zone \citep{HoC00}. 

\subsection*{The high frequency periods}
     \label{S-period-high} 
     
\begin{figure}[p]
\centering
\includegraphics[width=14.0cm,height=16.0cm]{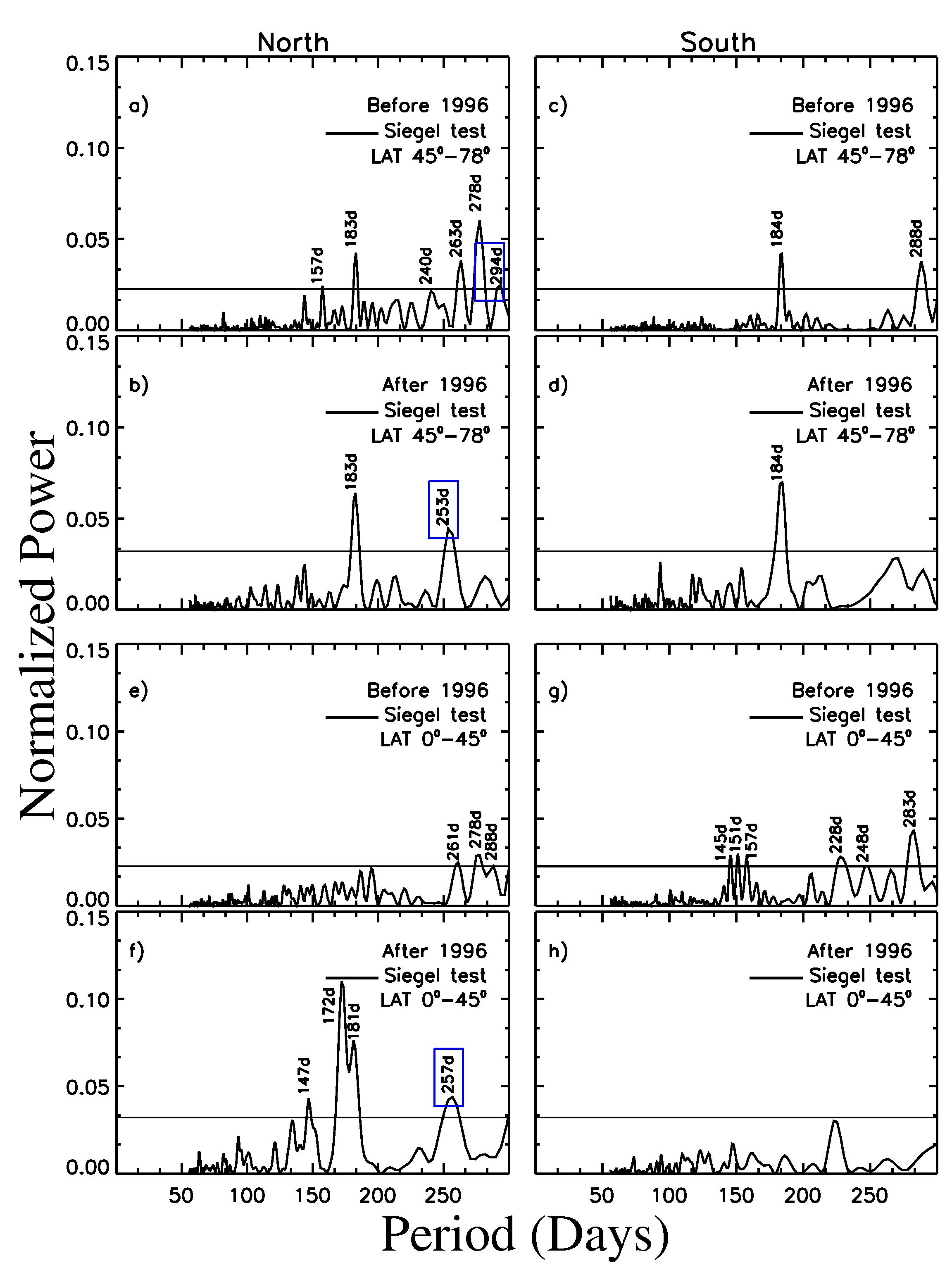}
\caption{The top four panels - a , b, c, d and the bottom four panels - e, f, g, h respectively 
represent the normalized Fourier periodograms for high-latitude fields and low-latitude fields for 
the high frequency periods in the frequency range 214 nHz to 38.5 nHz .  The respective four panels 
in the top and the bottom show the normalized Fourier power distribution with periodicity for 
photospheric fields in the north before 1996, in the north after 1996, in the south before 1996 and 
in the south after 1996. The new periodicities in our analysis have been boxed in blue.}
\label{F-asym-fields-high}
\end{figure}

In the high frequency range we have frequencies ranging from 214 nHz (54 d) to 
38.5 nHz (300 d).  These periodicities for both the high-latitude and low-latitude 
fields are listed in the Table \ref{T-tab-high}.  The upper half of the Table \ref{T-tab-high} 
lists periodicities for the high-latitude fields with periods ranging between 
157 d to 294 d.  Few of these periodicities with their Fourier spectral power are 
shown in the top four panels - a, b, c, d in Fig \ref{F-asym-fields-high}.  Similarly, the lower half 
of the Table \ref{T-tab-high} lists significant periodicities for the low-latitude fields 
with periods ranging in between 134 d to 288 d.  The bottom four panels - e, f, g, h 
in Fig \ref{F-asym-fields-high} show a few of these periodicities with their spectral power. 
As stated earlier the significant periodic components were determined using the Siegel 
test statistics and are those periodic components having power levels above the black horizontal 
line drawn in each panel of Fig. \ref{F-asym-fields-high}.

The known periodicities in the high-latitude zone include those with periods 
157 d, 183 d, 184 d, 240 d, 263 d, 278 d, and 288 d and the new periodicities ${\it{viz.}}$ 
with period 253 d and 294 d, not reported in the literature earlier, are shown in boldface 
in the upper part of the Table \ref{T-tab-high}. 

The Rieger periodicity of 157 d is a well known fundamental periodicity that we find only in 
the high-latitude fields in the northern hemisphere prior to 1996.  In addition, the periodicity 
of 182$-$184 d is seen to be always present and the Fourier power corresponding to this periodicity 
does not vary much in both the north and south fields at high latitudes both before and after 1996.  
These high-frequency periods, in the high-latitude fields, don't show much variation in their 
normalized Fourier power and we do not see any north-south asymmetry in these high-latitude fields. 

On the other hand, in the low-latitude zone, shown in the lower part of the Table \ref{T-tab-high}, 
we report periodicities of 145 d, 147 d, 151 d, 157 d, 172 d, 228 d, 248 d, 261 d, 278 d, 283 d and 288 d, 
which have already been reported in earlier work \citep{OBB98,KSo02, KSB05,CKR09}.  The new periodicity 
from our analysis was 257 d and is shown in boldface in the lower half of the Table \ref{T-tab-high}.  
The Rieger periodicity of 157 d is only seen in the south low-latitude field before 1996.  Also, the 
semi-annual variation of period 181 d is observed only in north low-latitude fields.  

\section*{Discussion and Conclusion}
 \label{S-Con}
Our wavelet analysis shows a transition occurring around 1996 in the distribution of power 
and periodicities of photospheric fields. When the data were partitioned into periods prior 
to and after 1996 a hemispheric asymmetry is seen in the derived Fourier periodicities 
of solar magnetic activity above latitudes of $\pm$45${^{\circ}}$. A more detailed analysis 
shows that this asymmetry is confined to the latitude band 45${^{\circ}}$$-$60${^{\circ}}$ in 
both hemispheres.  This latitude band is primarily dominated by strong, episodic poleward 
surges or tongues of magnetic flux in both hemispheres which, as stated earlier, are a direct 
surface manifestation of the meridional flow and the internal solar dynamo.  In the 
60${^{\circ}}$$-$75${^{\circ}}$ latitude band we find an asymmetry in the distribution 
of spectral power in the longer periods since the photospheric surface fields have saturated 
in this band. This observed and localized asymmetry, confined to the 45${^{\circ}}$$-$60${^{\circ}}$ 
latitude zone in both hemispheres, when coupled with the fact that both solar fields and the 
micro-turbulence levels in the inner-heliosphere have been decreasing since early to mid 
nineties suggests that active changes occurred around this time in the solar dynamo that governs 
the underlying basic processes in the sun. These changes in turn probably initiated the build-up 
to one of the deepest solar minima, at the end of the cycle 23, experienced in the past 100 years. 
%%------------------------------------begin Figure 7----------------------------------------
%\protect\begin{figure}[ht]
%\vspace{12.6cm}
%\center
%\special{psfile=Bisoi-etal-f07-new.eps
%angle=0 hscale=38 vscale=30 voffset=0 hoffset=70}
%%----------------------------------end Figure 7-------------------------------------------

The magnetic time series in both the solar hemispheres exhibit a multitude of periodicities with 
significant variation in the spectral power of midterm (1$-$2 year) periodicities before and after 
1996. These prominent periods in the lower solar latitudes, below ${\pm}$45${^{\circ}}$, are 
thought to originate as the result of stochastic processes caused by the periodic emergence of 
surface magnetic flux \citep{WSh03} as the solar cycle progresses.  As stated earlier, \cite{HoC00} 
have reported a 1.3 year periodicity at the base of the solar convection zone, which has also been 
detected in sunspot areas and sunspot number time series studied using wavelet transforms \citep{KSo02,CDw11}. 
These findings have led to the conclusion that the mid-term fluctuations in the solar fields are surface 
manifestations of changes in the magnetic fluxes generated deep inside the Sun. 
%%------------------------------------begin Figure 8---------------------------------------
%\protect\begin{figure}[ht]
%\vspace{11.0cm}
%\center
%\special{psfile=Bisoi-etal-f08.eps
%angle=0 hscale=55 vscale=55 voffset=0 hoffset=70}
\begin{figure}[p]
\centering
\includegraphics[width=12.0cm,height=12.0cm]{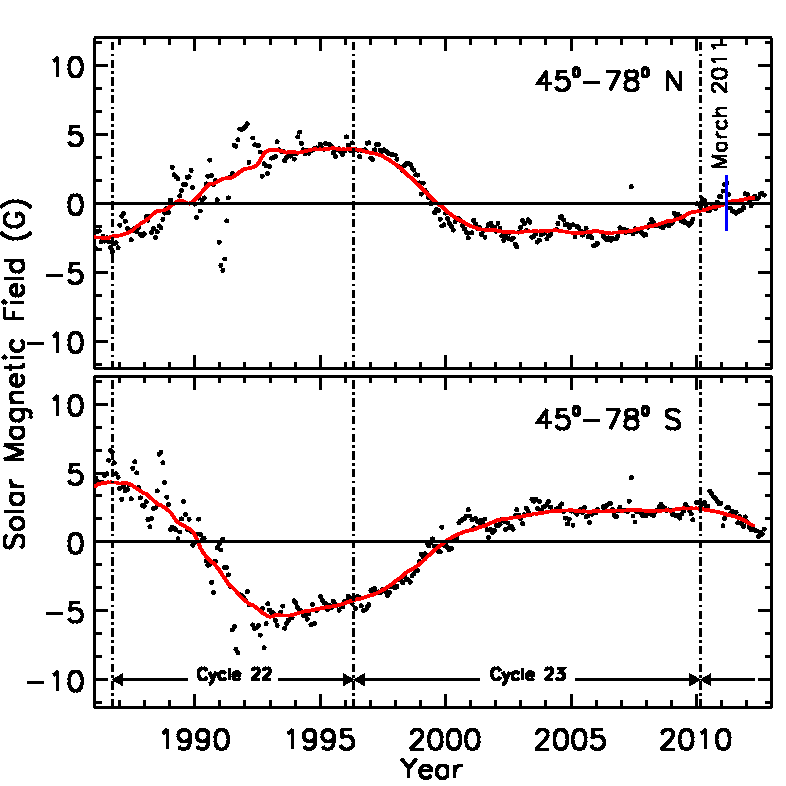}
\caption{Shows the solar magnetic field in the northern (top) and southern (bottom) hemispheres in the 
latitude range 45${^{\circ}}$ to 78${^{\circ}}$ for solar cycles 22 and 23.  It is clear that the north 
solar field has reversed while the south solar field is still to do so. The time of reversal of the north 
solar field is indicated by a small blue arrow.}
\label{F-reversal-assym}
\end{figure}
%%----------------------------------end Figure 8-------------------------------------------

Polar surges above latitudes ${\pm}$45${^{\circ}}$ show significant variations in their occurrence rates 
and strength in each cycle and will therefore show up as a variation in spectral power and periods. 
A comparison of polar surges in the last three solar cycles, {\textit{viz.}} 21, 22, and 23, reveals that 
they are comparatively more in strength after around solar maximum.  

Though we have restricted our analysis to data from cycles 21, 22 and 23 in the period from 
$\sim$1975~$-$~2010, an examination of data beyond 2010 shows an asymmetry in the solar field 
reversal at high-latitudes.  Figure \ref{F-reversal-assym} shows the reversal of the northern 
high latitude field in cycle 24 in March 2011.  However, the southern hemisphere is yet to 
undergo a reversal. The filled black dots in Fig. \ref{F-reversal-assym} are the actual 
measurements while the red solid line is a smoothed curve.  The reversal of the solar polar 
field occurs at times when the red curve passes through zero and is indicated for cycle 24 
by a small blue line in the northern hemisphere.  While it is known from earlier cycles that 
the two hemispheres do not reverse polarities at the same time, the time lag between the 
reversal of the northern hemisphere and southern hemisphere has, to the best of our knowledge, 
never been as large.  

In addition, \cite{DiG10} and \cite{Dik11}, using the theory of meridional 
circulation, have reported an asymmetry in the latitudinal extent of the Sun's meridional 
flow belt in the cycle 22 and 23 wherein surface meridional flows in cycle 22 extend to 
latitudes of  ${\pm}$60${^{\circ}}$ while in cycle 23, it went all the way to the poles. 
Thus, the meridional flow in cycle 23, took a longer time thereby causing a slower return 
flow which led in turn to the extended solar minimum in cycle 23. Recent work, using 
helioseismic data from the Birmingham Solar Oscillation Network (BiSON), an instrument 
that is very sensitive in probing the solar interior close to the solar surface, has shown 
that the behavior of solar oscillation frequencies in the sub-surface magnetic layers during 
the descending phase of cycle 23 was significantly different from that during cycle 22 
\citep{BaB12}.  These authors went on to state that the peculiar solar minimum, at the end 
of the cycle 23, could have been predicted long before it happened.  The present work, 
showing a north-south asymmetry around mid 1990's in the quasi-periodic variations of 
photospheric fields at high-latitudes shows that changes were initiated at this time in 
the basic underlying solar dynamo processes like the meridional flow rates and the magnetic 
flux emergence that eventually led to the prolonged and deep minimum that we have just 
witnessed.  

The current understanding of the solar dynamo is that it operates through the Babcock-Leighton 
mechanism to produce poloidal fields through the decay of tilted bipolar sunspots \citep{CCJ07,JCC07}.  
The strength of the polar field, at each solar maximum then depends upon the tilt angle in
bipolar sunspots which, in turn, is determined by the action of the Coriolis force acting on 
magnetic flux tubes that reach the surface by rising through the turbulent convection zone 
\citep{DCh93}.  This process causes a large scatter in the average tilt-angle causing the polar 
field to be randomly weaker or stronger than in the previous cycle \citep{LCh02}.  We have observed 
a decline in fields above 45${^{\circ}}$ for the past $\sim$15 years implying weak polar 
fields being generated in two successive solar cycles \textit{viz.}~cycles 22 and 23.  A 
continuation of this declining trend beyond 22 years would imply a third successive weak polar 
field imparted by the Babcock-Leighton mechanism.  This is probably unlikely as the polar field 
is imparted by a random process and if it happens it would have serious implications on our 
present understanding of the solar dynamo.  Also, two of the eight strongest geomagnetic 
storms in the last 150 years have occurred during solar cycles 13 and 14 which were both relatively 
weak cycles.  Therefore, continued observations and measurements of solar magnetic fields are 
extremely important.
 
\section*{acknowledgments}
{\it{The authors would thank the free use data policy of the National solar observatory and 
acknowledge for providing the data in the public domain via the World Wide Web. Wavelet 
software was provided by C. Torrence and G. Compo, and is available at 
http://paos.colorado.edu/research/wavelets.}}

%
%\mbox{}~\\ 
%\bibliographystyle{spr-mp-sola}
%\bibliography{jdm_swd}  
%

%\end{article} 
%
\end{document}